\documentclass[journal]{IEEEtran}
\usepackage[pdftex]{graphicx}
\usepackage{amsmath}
\usepackage{amsfonts}
\usepackage{bbm}

\usepackage{threeparttable}
\usepackage{booktabs}
\usepackage{cite}
\usepackage{makecell}

\graphicspath{{fig/}}
\begin{document}
\title{Improving Sample Efficiency of Deep Learning Models in Electricity Market
}

\author{
	Guangchun~Ruan,
	Jianxiao~Wang,
	Haiwang~Zhong,
	Qing~Xia,
	and Chongqing~Kang
}

\maketitle

\begin{abstract} 
The superior performance of deep learning relies heavily on a large collection of sample data, but the data insufficiency problem turns out to be relatively common in global electricity markets. How to prevent overfitting in this case becomes a fundamental challenge when training deep learning models in different market applications.
With this in mind, we propose a general framework, namely Knowledge-Augmented Training (KAT), to improve the sample efficiency, and the main idea is to incorporate domain knowledge into the training procedures of deep learning models. Specifically, we propose a novel data augmentation technique to generate some synthetic data, which are later processed by an improved training strategy. This KAT methodology follows and realizes the idea of combining analytical and deep learning models together. Modern learning theories demonstrate the effectiveness of our method in terms of effective prediction error feedbacks, a reliable loss function, and rich gradient noises.
At last, we study two popular applications in detail: user modeling and probabilistic price forecasting. The proposed method outperforms other competitors in all numerical tests, and the underlying reasons are explained by further statistical and visualization results.
\end{abstract} 


\section{Introduction} \label{sec:intro}
\subsection{Background}
Deep learning has achieved remarkable success in the past decades, which highly depends on the increasing data volume and computing power~\cite{goodfellow2016deep}. Collecting a large amount of data in many real-world applications, e.g., user analysis in electricity markets, is unfortunately difficult, expensive, or even forbidden. This raises a common problem of data insufficiency~\cite{wang2020generalizing}, and we list four reasons below to further explain this issue in the context of electricity market:
\begin{itemize}
	\item Moderate or low sampling frequency is widely applied in electricity markets. For instance, most data in day-ahead markets are recorded on a daily basis, so that one can only collect around 360 data sample each year.
	\item Many market data are very sensitive due to privacy concerns, especially trade-related or financial data. The availability of these data is thus strictly limited.
	\item Market conditions and user preferences are changing rapidly, so only the recent data can provide effective information. Some bidder, for example, may change his bidding strategy within a few months.  
	\item Complex market structure and irrational user behaviors call for a large amount of descriptive data to enable fine-grained modeling.
\end{itemize}

There is an apparent contradiction between the data requirement for high-capacity models and the limited amount of available data. In general, training a sophisticated deep learning model with limited amount of data is risky and unreliable due to the overfitting problem~\cite{mohri2018foundations}. A promising solution is to enhance the sample efficiency, i.e., to extract as much effective information as possible from a small dataset. The domain knowledge may offer special opportunities in this aspect---we already have a library of classical models, which are rigorously developed and validated in different electricity markets, and thus provide a good starting point for the deep learning models.

Motivated by the above observations, this paper intends to make full use of these existing classical models when training deep learning models. Our main focus is on the sample efficiency of regression and supervised learning tasks, and the proposed methodology should have a broad applicability in different applications.

\subsection{Literature Review}
A large number of publications have used deep learning models in different market applications, e.g., price forecasting~\cite{lago2018forecasting}, load forecasting~\cite{yan2018multistep,zeng2019dynamic,bottieau2020veryshortterm}, peer-to-peer trading~\cite{qiu2021scalable}, demand response~\cite{kim2020supervisedlearning}, and bidding~\cite{ruan2021bidcurv}. 
Much more sophisticated models are surging in recent years, but little attention has been paid to the data insufficiency issue. Some of existing works tend to partly use the simulation data instead~\cite{kim2020supervisedlearning}, some makes implicit assumptions that market conditions keep unchanged for five years or beyond~\cite{lago2018forecasting}. In practice, however, it is crucial for market participants to work effectively with those real, dynamic, imperfect, and often insufficient data.

The machine learning community has discussed this issue in the regime of few-shot learning~\cite{du2020fewshot}. The key point is how to match the available data volume with the data required for model calibration. Many few-shot learning models are based on some kinds of similarity metrics, including MatchingNet~\cite{vinyals2016matching}, PrototypeNet~\cite{snell2017prototypical}, and RelationNet~\cite{sung2018learning}. 
Another popular pathway is using extra information or data sources, e.g., side information~\cite{visotsky2019fewshot}, or knowledge transfer~\cite{finn2017modelagnostic,munkhdalai2017meta} (also known as transfer learning). But cross-market datasets are often hard to use because of the inconsistent data structure and precision.
The community has developed a special technique called data augmentation to systematically enlarge the datasets. Although there are several adaptive variants such as generative adversarial networks~\cite{antoniou2017data,cubuk2018autoaugment}, the mainstream idea for data augmentation is simply using geometric transformation~\cite{taylor2017improving}, e.g., flipping, rotating, cropping, and color jittering.

These efforts from the machine learning community may deepen our understanding of the problem, but their tasks are quite different from ours in essence---we have better knowledge of how electricity markets are operating, but our data volume is much smaller. In other word, the market applications are often knowledge-rich but data-limited. This finding also partly demonstrates why the existing strategies from the machine learning community perform poorly in electricity market applications.

Hybrid analytical and data-driven modeling is another related topic~\cite{ruan2020mlopt}. The most common idea is replacing the inaccurate analytical parts with a neural network, so that the hybrid systems are operating either in a coupled manner~\cite{kim2020supervisedlearning,gutierrez2011neuralnetwork} or an iterative manner~\cite{ruan2020nnlms}. 
In addition, the physics-guided neural networks or physics-informed neural networks have been established in the existing literature, but there is no general rule for such designs, and an inefficient scheme may severely degrade the model performance.

Reference~\cite{rai2020driven} has reviewed four patterns for hybrid modeling, i.e., physics-based preprocessing, physics-based network architectures, physics-based regularization, and miscellaneous architectures. Within this taxonomy, reference~\cite{wang2019exploring} preprocessed the input data with an analytical model to derive the irradiance components and cell temperatures, reference~\cite{zhou2017partial} improved the loss function in a semi-supervised learning task, reference~\cite{qiu2022safe} constructed the regularization scheme in safe reinforcement learning by adding penalty on potential violations of physical constraints, and reference~\cite{yang2020fast} designed a specific network structure with the guidance of power flow equations. 
Special network structures have also been studied to reflect some problem-related features. A convex neural network was developed in \cite{chen2018optimal} for optimal control, and a graph convolutional neural network was employed in optimal load shedding~\cite{kim2019graph} and wind power estimation~\cite{park2019physicsinduced}. 
Furthermore, differential equations were used to formulate the neural networks in \cite{misyris2020physicsinformed}, while the inaccurate analytical model offered a basic policy for a learning machine in \cite{zhao2020hybrid}.

Note that this area of hybrid modeling is still in a nascent stage, which lacks systematic testing and in-depth explanations of the method effectiveness.

Perhaps the most exciting opportunity in electricity market is that experts have developed a lot of classical models, which fully reflect the domain knowledge. For example, most markets are prevalently operated in a strict optimization-based formulation, which provides side information to boost learning performance. This paper makes the first attempt to utilize these candidate models to address the data insufficiency issue and thus achieve a higher sample efficiency.

\subsection{Contributions and Paper Structure}
The major contributions of this paper are listed as follows:
\begin{enumerate}
	\item A general framework, Knowledge-Augmented Training (KAT), is proposed to improve the sample efficiency and avoid overfitting of deep learning models. This framework can make full use of domain knowledge to guide the training procedures of neural networks. Different from the existing hybrid modeling, this framework elaborates a special and flexible coordination between analytical and deep learning models.
	
	\item A novel data augmentation technique and an improved training strategy are developed within the proposed framework. Combining these two methods makes it possible to achieve high learning performance while keeping low sample complexity. Technically, this procedure involve the designs of calibration, filtering, aggregation, and dynamic sampling.
	
	\item The method effectiveness is explained by the modern learning theories as well as several statistical and visualization results from two typical market applications.
\end{enumerate}

The rest of this paper is organized as follows. Section~\ref{sec:framework} introduces the overall framework, in which a data augmentation technique and an improved training strategy are further elaborated in Section~\ref{sec:augmentation} and \ref{sec:training} respectively. We then demonstrate the method effectiveness in Section~\ref{sec:effectiveness}. Section~\ref{sec:case} analyzes two typical market applications, and at last, Section~\ref{sec:conclusion} draws the conclusion. We include the nomenclatures and acronyms in the appendix.

\section{Framework} \label{sec:framework}

\subsection{Problem Statement}
Sample efficiency is a major concern for deep learning models when their training datasets are insufficient, and this paper is focused on such issue in the context of electricity market. We use a small historical dataset $\mathit{HD} = \{ (x_1, y_1),\cdots,(x_n, y_n) \}$ to train a deep learning model $y=f(x;\theta)$, here $\theta$ denotes the model parameters. The data insufficiency suggests that in case $\dim \theta \gg n$, the deep learning model $f(\cdot;\theta)$ will easily overfit and become problematic.

A potential resource in electricity markets is the classical models established by domain experts. We show these models by $y=g_k(x;w_k),\ k=1,2,\cdots,K$, here $w_k$ denotes the model parameters, and usually $\dim w < n$.

Formally, the objective of this paper is to improve the sample efficiency, or avoid overfitting, of a deep learning model $f(\cdot;\theta)$ when a small historical dataset $\mathit{HD}$ and some classical models $g_k(\cdot;w_k)$ are given.

Note that the details of nomenclatures and acronyms are presented in the appendix for ease of reference.

\begin{figure*}[t] 
	\centering 
	\includegraphics[width=1\textwidth]{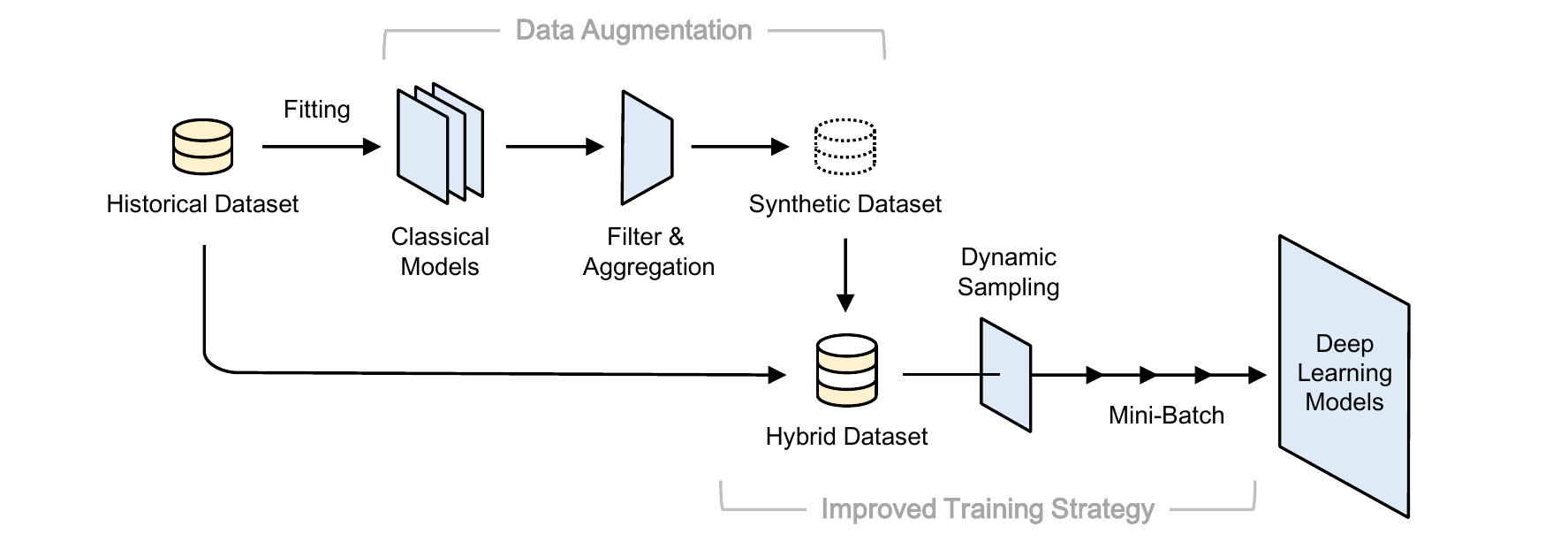}
	\caption{The proposed Knowledge-Augmented Training framework for the sample efficiency improvement. There are two technical parts: the data augmentation technique and improved training strategy. Data augmentation is used to generate some synthetic data, which are later incorporated into an improved training procedure. This framework is flexible and powerful to cover many practical cases.} 
	\label{fig:framework} 
\end{figure*} 

\subsection{Overall Framework}
Fig.~\ref{fig:framework} has shown a few key components of the proposed framework. The basic workflow is enlarging the data volume first by a data augmentation technique, and then calibrating the deep learning models with an improved training strategy.

A synthetic dataset $\mathit{SD} = \{ (x_{n+1}, y_{n+1}),\cdots,(x_N, y_N) \}$ is generated by running the classical models until $N > \dim \theta$. Note that we only distinguish these synthetic data and the historical data by indices. Next, we merge these two data sources and get a hybrid dataset $D = \mathit{HD} \cup \mathit{SD}$. To accelerate the training procedure, a mini-batch $D_m = \{ (x_{(1)}, y_{(1)}),\cdots,(x_{(m)}, y_{(m)}) \}$ is dynamically sampled from this hybrid dataset, so $D_m \subset D$, usually $m \ll N$. We will update the model parameters $\theta$ later with a gradient step calculated with $D_m$. 

\begin{figure}[t] 
	\centering 
	\includegraphics[width=0.48\textwidth]{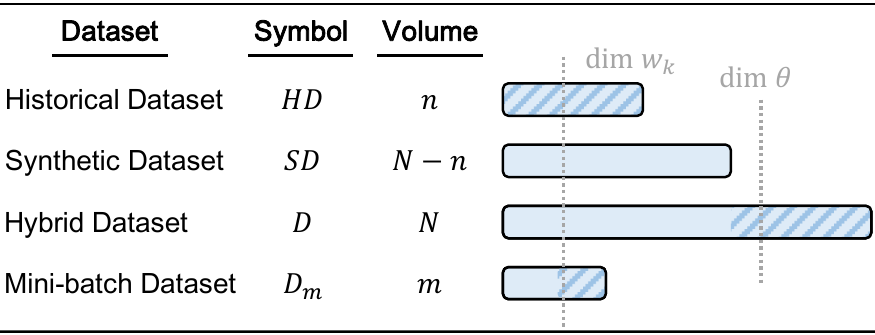} 
	\caption{Summary of the proposed datasets. The horizontal bar chart gives an illustrative example of the data volume, and the numbers of parameters in deep learning models and classical models are given for comparison.} 
	\label{fig:dataset} 
\end{figure}

Fig.~\ref{fig:dataset} provides a summary of the four datasets mentioned above, and more technical details can be found in Section~\ref{sec:augmentation} and \ref{sec:training}.

We summarize two distinct advantages of the proposed framework. First, this framework is able to efficiently combine the domain knowledge and learning approaches together to achieve better performance. Second, as shown later in Section~\ref{sec:case}, this methodology is flexible and may readily adapt to different practical applications.

Note that the punchline lies in a better design of coordinating the analytical models (domain knowledge) with learning models (data-driven), and one can instantiate the whole framework by replacing the placeholders with any specific models.

\section{Data Augmentation} \label{sec:augmentation}
Data augmentation is a popular technique to generate new (synthetic) data from existing historical data. In this section, we focus on a specific kind of data augmentation technique that makes full use of the domain knowledge.

\subsection{Calibration of Classical Models}
We first calibrate the classical models one by one to match the data observations, and these calibrated models might be useful in generating synthetic data later. Suppose $K$ different models are available, and the calibration problem for model~$k$ can be formulated as follows,
\begin{align} 
	\label{eqn:Pk} 
	e_k = \min_{w_k}\ \frac{1}{n} \sum_{i=1}^n \big( g_k(x_i; w_k) - y_i \big)^2,
\end{align}
where $e_k$ is the mean squared error between the model outputs $g_k(x_i; w_k)$ and actual observations $y_i$. Minimizing this error term will derive the optimal parameter $w_k^*$ along with the minimum $e_k^*$. Note that $e_k^*$ is a suitable metric to evaluate the applicability of model~$k$.

Now, our key focus is on the solution of~(\ref{eqn:Pk}). When $g_k(\cdot;w_k)$ is explicitly expressed by some function, we can apply the famous Broyden--Fletcher--Goldfarb--Shanno (BFGS) algorithm to solve (\ref{eqn:Pk}) efficiently. 

The problem raises when $g_k(\cdot;w_k)$ is not explicitly available. For example, $y=g_k(x;w_k)$ may be hidden within an optimization model where the optimal variable is $y$ and $x$ is an external input. This kind of problem is often called inverse optimization~\cite{aswani2018inverse}. In such a case, we can integrate the Karush--Kuhn--Tucker (KKT) conditions into the constraints, and reformulate (\ref{eqn:Pk}) as follows,
\begin{subequations} 
	\label{eqn:CMk} 
	\begin{alignat}{2}
	\min_{w_k, \tilde{y}_i, \lambda_i} 
	&\ \frac{1}{n} \sum_{i=1}^n \big( \tilde{y}_i - y_i \big)^2, \\
	\textrm{s.t.} \ \ 
	&\ h^\textrm{KKT} (x_i, \tilde{y}_i, w_k, \lambda_i) \le 0, \ \forall i,
	\end{alignat}
\end{subequations}
where $h^\textrm{KKT}(\cdot)$ is a compact way to express a group of KKT conditions. $\lambda_i$ is the Lagrange multiplier, and $\tilde{y}_i$ is an intermediate variable for model descriptions. 

We further use the big-M method to linearize the above KKT constraints, and derive a mixed-integer linear programming (MILP) model. Such models can be effectively solved by the Branch-and-Bound (BnB) algorithm. 

When dealing with some too-complicated models, evolutionary methods such as Particle Swarm Optimization~(PSO) and Genetic Algorithm~(GA) can be adopted instead.

It should be highlighted that this calibration step could be bypassed in case some calibrated models are already ready. This is very common in the situation of model upgrade. Many market participants are interested in developing deep learning models to replace their time-series models in history.

\subsection{Filter and Aggregation}
After calibrating all the candidate models, the next step is to aggregate the model outcomes in order to generate the synthetic data. We enhance the quality of synthetic data by selecting the most accurate models and averaging their outputs to avoid individual biases.

Given a series of $x_i,\ i \ge n+1$, the synthetic data $(x_i, y_i)$ can be generated with the following formula,
\begin{subequations} 
	\label{eqn:ensemble} 
	\begin{alignat}{2}
	& \gamma_k = \frac{(e_k^* + \alpha)^{-1} I(e_k^* < \beta)}
	{\sum_{k=1}^K (e_k^* + \alpha)^{-1}  I(e_k^* < \beta)}, \ \forall k, \\
	& y_i = \sum_{k=1}^K \gamma_k \, g_k(x_i; w_k^*), \ \forall i \ge n+1,
	\end{alignat}
\end{subequations}
where $\gamma_k$ is a normalized weighting factor that is inversely proportional to the model errors $e_k^*$. $I(\cdot)$ is an indicator function that outputs 1 if the inner condition is satisfied, and outputs 0 otherwise. $\alpha$ and $\beta$ are two user-defined constants.

Equations~(\ref{eqn:ensemble}) have shown our strategy to derive reliable synthetic data---filtering out those models that cannot reach the accuracy requirement $\beta$, and then averaging the eligible outputs with some tailored weights $\gamma_k$.

\subsection{Other Configurations}
We briefly discuss the settings of $N$ and $x_{n+1},\cdots,x_N$ here. Often, $N$ is set as a large number such that $N > \dim \theta$. There are many possible ways to choose $x_{n+1},\cdots,x_N$, but we recommend reducing the intersection between these data and the historical ones. In some use cases, we can sample $x_{n+1},\cdots,x_N$ from an interval or a region with rare historical observations. This is helpful in practice to provide new information that is originally unknown.

\section{Improved Training Strategy} \label{sec:training}
As synthetic data are available, a novel strategy should be carefully designed to make full use of these data resources while reducing the potential influences of data noises.

\subsection{Mini-Batch Gradient Descent}
For regression tasks, deep learning models are often trained and calibrated by minimizing the loss function:
\begin{align} 
	\label{eqn:JN} 
	J_N(\theta)=\frac{1}{N} \sum_{i=1}^N F \big( f(x_i;\theta) - y_i \big).
\end{align}

Different loss functions $J_N(\cdot)$ can be derived when using different formulations of $F(\cdot)$, e.g., the well-known mean squared error and the pinball loss. We consider the mean squared error by default: $F(y^\text{pred}_i, y_i) = (y^\text{pred}_i - y_i)^2$.

Using the whole dataset for training is found inefficient in terms of computing speed and training effectiveness~\cite{goodfellow2016deep}. Mini-batch training is established to overcome this issue by taking a subset instead of the whole dataset during one training iteration~\cite{li2014efficient}. The following expression shows how to calculate the gradient on a mini-batch:
\begin{align} 
	\label{eqn:dJm} 
	\nabla J_m(\theta)=\frac{2}{m} \sum_{i=1}^m \big( f(x_{(i)};\theta) - y_{(i)} \big) \nabla_\theta f(x_{(i)};\theta),
\end{align}
where $\nabla_\theta f(x_{(i)};\theta)$ is obtained by the Backpropagation (BP) algorithm. This gradient will be updated at each iteration after mini-batch resampling (multi-fold or randomly).

The model parameters $\theta$ are randomly initialized, and we update these parameters with a series of gradient steps, 
\begin{align} 
	\label{eqn:grad} 
	\theta \leftarrow \theta - s \nabla J_m(\theta),
\end{align}
where $s$ is the step size. By default, we choose the Adam algorithm for training.

\subsection{Dynamic Sampling}
Different from classical mini-batch methods, the proposed methodology is special in the data source---a combination of both historical and synthetic data. Since synthetic data are expected to be noisy, it is important to analyze the data composition of each mini-batch.

We split a mini-batch into two parts, and get $m = m_1 + m_2$. Here $m_1 = | D_m \cap \mathit{HD} |$ shows how many data are sampled from historical dataset, and $m_2 = | D_m \cap \mathit{SD} |$ shows the case for synthetic dataset. The symbol $|\cdot|$ is used to express the cardinality (or size) of a given set. 

We further introduce a metric to quantify the proportion of historical data as follows:
\begin{align} 
	\label{eqn:rate} 
	\eta = \frac{m_1}{m} \times 100\% \ \in \ \big[\eta_\textrm{min}, \eta_\textrm{max}\big],
\end{align}
where $\eta_\textrm{min}$ and $\eta_\textrm{max}$ are the restricted lower and upper bounds that are configured by users. A reasonable lower bound is $\lfloor mn/N \rfloor$ ($\lfloor \cdot \rfloor$ is a floor function, this bound is derived by uniform sampling), and the upper bound should not exceed 90\% as recommended.

The basic idea of dynamic sampling is to ensure $\eta$ gradually increases when the training goes on. Given a large number $T$, we can initialize $\eta=\eta_\textrm{min}$, and update it by calculating:
\begin{align} 
	\label{eqn:rate-update} 
	\eta \leftarrow \eta + \frac{1}{T} \big(\eta_\textrm{max} - \eta_\textrm{min}\big).
\end{align}
Here $\eta$ will keep unchanged after reaching the maximum $\eta_\textrm{max}$.

Using this ratio $\eta$ to construct a mini-batch is intuitive: randomly take $\lfloor m\eta \rfloor$ samples from the historical dataset $\mathit{HD}$, then take $m - \lfloor m\eta \rfloor$ samples from the synthetic dataset $\mathit{SD}$, and finally merge all samples together.

\begin{figure}[t] 
	\centering 
	\includegraphics[width=0.48\textwidth]{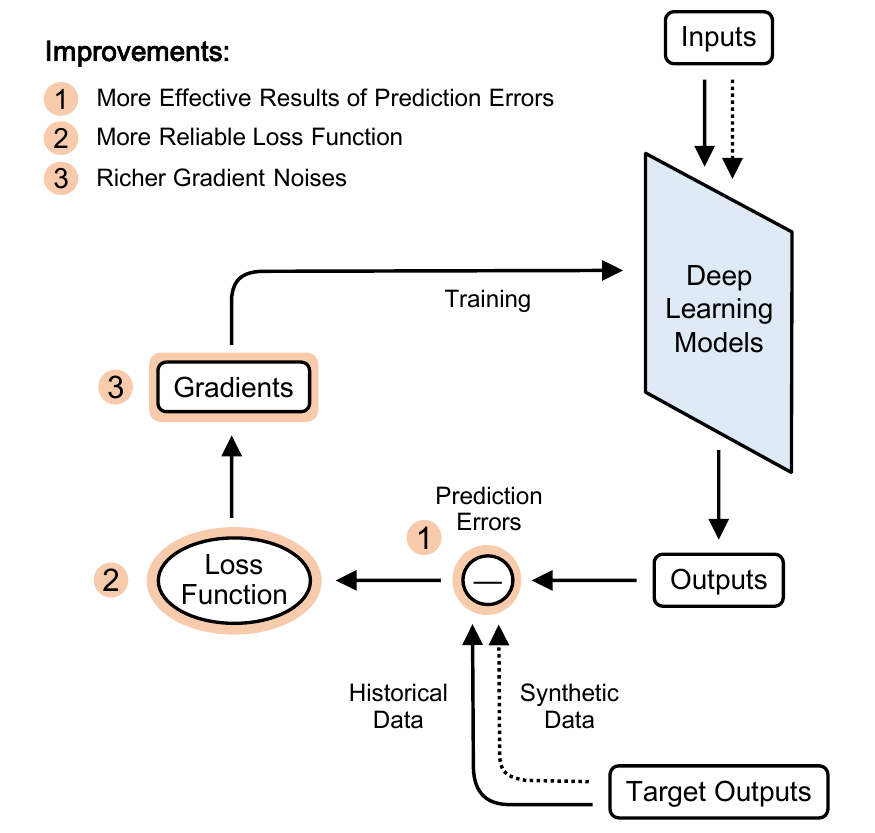}
	\caption{Explanations of the method effectiveness. The proposed methodology is expected to achieve three improvements in the training procedure of deep learning models.} 
	\label{fig:effect} 
\end{figure} 

\section{Method Effectiveness} \label{sec:effectiveness}
This section attempts to introduce some latest theoretical findings to explain why the proposed methodology could be effective. Fig.~\ref{fig:effect} highlights three improvements of our method from a training perspective, and more details are discussed in the following subsections.

\subsection{More Effective Feedbacks of Prediction Errors} \label{subsec:prediction-error}
A major concern about the proposed methodology is that synthetic data are not exactly accurate, and it is therefore uncertain whether these noisy data could improve the training performance of deep learning models. In fact, synthetic data are powerful to provide some approximated manifold structures that make it easier to train those models. 

\begin{figure}[t] 
	\centering 
	\includegraphics[width=0.48\textwidth]{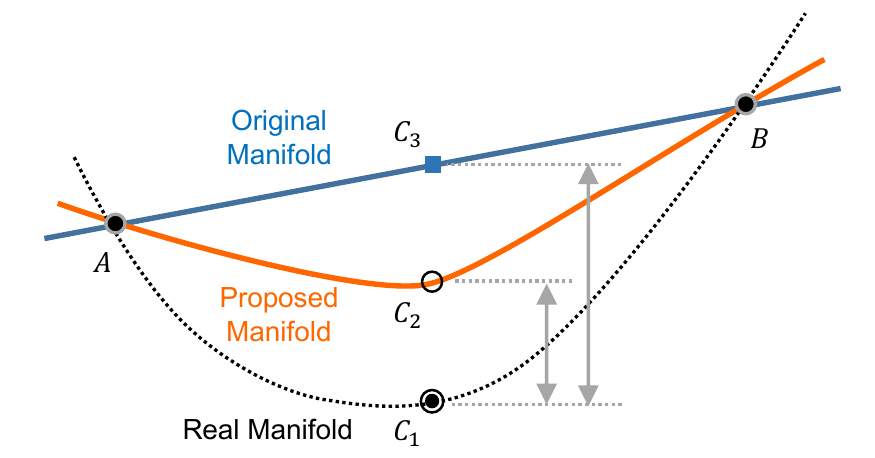} 
	\caption{Illustrative example of the prediction errors between different manifolds. The proposed methodology can provide approximated manifold structure to improve the learning results.} 
	\label{fig:manifold} 
\end{figure}

Fig.~\ref{fig:manifold} provides an illustrative example to understand the role of synthetic data by analyzing the error terms. Five points are labeled as A, B, C$_1$, C$_2$, and C$_3$. Here A, B~$\in \mathit{HD}$, C$_2 \in \mathit{SD}$, C$_3$ is a point on the original manifold, C$_1$ is the ground truth, and we assume C$_1$, C$_2$, and C$_3$ are vertically aligned. Comparing the original manifold and the proposed manifold will certainly explain why the synthetic data might be helpful.

Assume the coordinates of C$_1$, C$_2$, C$_3$ are $(x_i, y_i^\textrm{real})$, $(x_i, y_i)$, and $(x_i, f(x_i;\theta))$ ($i \ge n+1$). We thus analyze the manifold difference as follows,
\begin{align} 
	\label{eqn:error} 
	f(x_i;\theta) - y_i = (f(x_i;\theta) - y_i^\textrm{real}) - (y_i - y_i^\textrm{real}).
\end{align}

Consider the right-hand side of (\ref{eqn:error}). The first term describes the distance between C$_1$ and C$_3$, while the second term is the distance between C$_1$ and C$_2$. We can find in this example that although synthetic data are not perfect (the second term is larger than zero), they are practically useful and better than nothing (the first term is larger than the second term).

Such situation is not a single case, and synthetic data can practically enhance the training performance by providing more noisy but effective results of prediction errors.

\subsection{More Reliable Loss Function} \label{subsec:loss-func}
Another important focus of method effectiveness is on the generalization performance. Training strategy always centers on the generalization ability---we train a neural network on a sample set, but expect a superior out-of-sample performance. 

Generalization performance is heavily influenced by the selection of loss functions. As an empirical risk metric, an ideal loss function is expressed as follows~\cite{goodfellow2016deep},
\begin{align} 
	\label{eqn:J} 
	J(\theta)=\int \big( f(x;\theta) - y \big)^2 \textrm{d}p(x,y),
\end{align}
where $\textrm{d}p(x,y)$ denotes the ground truth joint distribution of data samples that are hardly available in practice. 

People often approximate this ideal function $J(\theta)$ by considering finite number of sample data. In theory, such approximation becomes less reliable when using smaller datasets. Data insufficiency issue thus leads to a unreliable loss function~\cite{wang2020generalizing}, typically $J_n(\theta)$, that may severely deteriorate the generalization performance of deep learning models.

Enlarging data volume by the synthetic data is a promising way to improve the above approximation. The data noise may bring some troubles, but we find the benefits are practically dominant. As a loss function, $J_N(\theta)$ is often more reliable than $J_n(\theta)$.

\subsection{Richer Gradient Noises} \label{subsec:grad-noise}
The final improvement lies in the training dynamics that can be analyzed by gradient noises. Here, we define the gradient noises as follows,
\begin{align} 
	\label{eqn:gn} 
	\mathit{GN}(\theta) = \nabla J_m(\theta) - \nabla J_N(\theta).
\end{align}

In modern learning theories, these gradient noises might increase the local search capability of deep learning models, and thus avoid falling into some local minima of high cost~\cite{goodfellow2016deep}. We are more likely to derive a model with stronger generalization performance if richer gradient noises appear in the training procedure~\cite{zhou2020towards}.

The randomness of mini-batch training is introduced by the data sampling. In this context, the proposed data augmentation technique provides more diverse samples, and the dynamic sampling strategy further generates richer data combinations. All these measures contribute to creating more noisy gradients.

\section{Case Study} \label{sec:case}
\subsection{Scenario Setup}
This section considers two typical market applications that could be improved by the proposed framework and methodology, they are user modeling (Case~1) and probabilistic price forecasting (Case~2). We pick these two cases mainly because they are popular, critical, and fundamental for quite a lot of decision-making tasks; they also collectively validate the applicability of our method in applications with different characteristics. In order to avoid too many repetitions, we will show the major procedure and key steps in Case~1, and supplement more statistical details in Case~2 to complete the full discussions.

Since there are a few possible options to deal with data insufficiency issue~\cite{goodfellow2016deep}, we could summarize these options from either a dataset or a model formulation perspective. Formally, we consider the following four kinds of datasets:
\begin{itemize}
	\item \textbf{O}: The original (historical) dataset $\mathit{HD}$.
	\item \textbf{KAT}: The hybrid dataset $D$ generated by the proposed data augmentation technique. Select $\eta_\textrm{min}=0.5, \eta_\textrm{max}=0.95, T=600$ to generate the synthetic data batches.
	\item \textbf{DA1}: The enlarged dataset generated by a sampling-and-averaging approach. Randomly sample three records from $\mathit{HD}$ and takes the average value as a new data.
	\item \textbf{DA2}: The enlarged dataset generated by adding random noises. Randomly sample one record from $\mathit{HD}$ and adds $\pm 0.5\%$ fluctuating noises to generate a new data.
\end{itemize}

Note that DA1 and DA2 are two commonly-used data augmentation approaches for regression tasks, and we will show later that KAT work better than both of them.

In addition, we pay attention to the following three kinds of model formulations:
\begin{itemize}
	\item \textbf{LM}: The large-scale deep learning model whose detailed formulations and hyper-parameters will be specified in the subsections later.
	\item \textbf{SM}: The small-scale deep learning model, specified later.
	\item \textbf{RM}: The regularized deep learning model that has a large-scale formulation and dropout layers.
\end{itemize}

For all cases below, we train the above deep learning models for 4000 iterations by the Adam algorithm.

Combining the above symbols together, we derive six scenarios of interest:
\begin{itemize}
	\item \textbf{KAT-LM}: Training the large-scale deep learning model LM with the hybrid dataset. Following the KAT framework, the proposed data augmentation and improved training strategy are used to boost the performance.
	\item \textbf{O-LM}: Training the large-scale deep learning model LM with the original dataset O (or $\mathit{HD}$).
	\item \textbf{O-SM}: Training the small-scale deep learning model SM with the original dataset O.
	\item \textbf{O-RM}: Training the regularized deep learning model RM with the original dataset O.
	\item \textbf{DA1-LM}: Training the large-scale deep learning model LM with the enlarged dataset DA1.
	\item \textbf{DA2-LM}: Training the large-scale deep learning model LM with the enlarged dataset DA2.
\end{itemize}

Note that the above six scenarios have covered the well-known options that could probably handle the data insufficiency issue, including limiting the model size (O-SM), using regularization (O-RM), and enlarging the original dataset (DA1-LM, DA2-LM). 

We discuss the above scenarios of interest in both Case~1 and Case~2, and the specific data and model formulations are given when necessary. We further distinguish the scenarios in Case~1 and Case~2 by different labels at the end, for instance, KAT-LM[UM] for the user modeling in Case~1 and KAT-LM[PPF] for the probabilistic price forecasting in Case~2.

On the basis of Case~1 and Case~2, we further provide two extended tests to conduct sensitivity analysis and consider another two hybrid models: \textbf{CT-NN}, a clustering-based neural network derived from~\cite{panapakidis2016day}; and \textbf{PINN}, a physics-informed neural network inspired from~\cite{bigerna2018estimating}. More details are available in the corresponding subsection.

Results in the following subsections are all simulated on a laptop with Intel i7-8550U CPU and 16.0 GB RAM. The programming environment is Python 3.6.0 with Tensorflow 1.12.0 (Keras backend), Gurobipy 8.0.0, and PySwarm 1.3.0. We follow the common practice to train each model for 10 times and then calculate the average performance as well as the uncertainty intervals.

\subsection{Case 1: User Modeling}

\subsubsection{Description}
User modeling refers to describing the responsive electricity consumption of a user under different incentives. We concentrate on the price responsive behavior that is frequently discussed in the existing literature. 

We collect the day-ahead price data in 2018 and 2019 from PJM market. The electricity user is assumed to be an industrial factory with several time-shiftable loads, and its dispatch model derived from~\cite{ruan2021bidcurv} is a high-dimensional mixed integer linear programming formulation. This model is simulated to generate the responsive electricity consumption data. The data from January 1, 2018 to June 30, 2019 are used as training data, while the rest are test data. 
We only consider the data in two years because users are always dynamic and tend to change their energy consumption behaviors shortly. In this case, only 546 training data are available, which cannot reliably train large-scale models.

Three data augmentation techniques (used in KAT, DA1, and DA2) will generate 1500 data and increase the total data volume to 2046. The classical model within the KAT framework is a general optimization model with constraints for upper and lower bounds, ramping rates, and total consumption~\cite{wang2017distributed}---Such a model turns out to be efficient and widely used in practice. Following~(\ref{eqn:CMk}), we calibrate this model by the PSO algorithm and achieve a 77.6\% accuracy. Note that it is possible in practice to choose a more dedicated model according to the domain knowledge.

We focus on the application of dense neural networks (NN) in user modeling. The desired neural networks should have an input of day-ahead prices and an output of responsive electricity consumption. The mean squared error is chosen as the loss function, and ReLU as the activation function. The detailed structures of LM, SM, RM are configured as 24-48-24, 24-9-9-24, and 24-48-24 plus 0.5 dropout rate, respectively.

Combining all the above data and models together will demonstrate the configuration details of six scenarios of interest. We next turn to the numerical tests and analysis part.

\begin{figure}[t]
	\centering
	\includegraphics[width=0.49\textwidth]{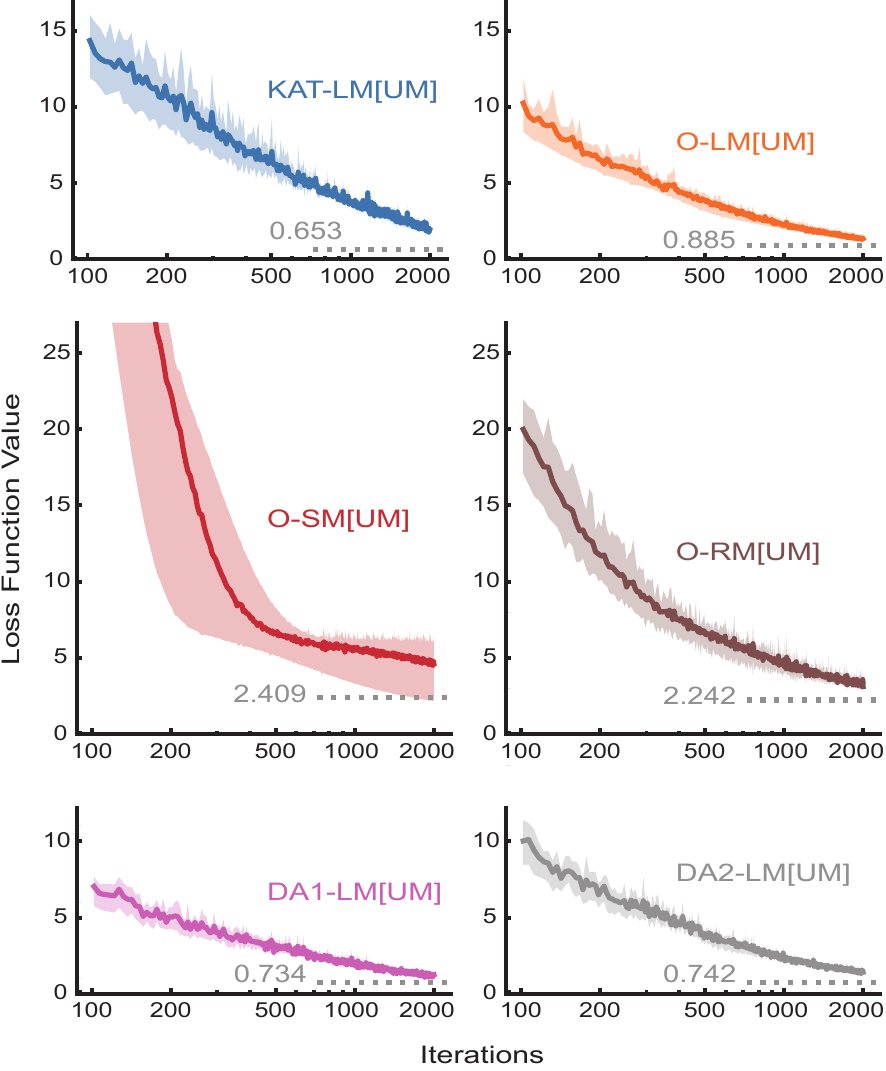}
	\caption{Comparison of the training procedures of different scenarios. The loss series of each scenario is visualized with a fragment of early stage (10--2000 iterations), and the final convergence level after 4000 iterations is given by the dashed gray line.}
	\label{fig:loss-series}
\end{figure}

\subsubsection{Accuracy Performance}
Fig.~\ref{fig:loss-series} compares the training procedures of different scenarios by plotting the loss series between 10--200 iterations. With a logarithmic-scale x-axis, one can clearly find a first-order convergence rate in KAT-LM[UM], O-LM[UM], DA1-LM[UM], and DA2-LM[UM], while the rest two scenarios converge much slower. This is mainly because limiting the model size or regularization could influence the fitting power and flexibility. Fig.~\ref{fig:loss-series} also gives the convergence values after 4000 iterations, and KAT-LM[UM] has the best convergence status of 0.653, which is 53.4\% lower than the average of other scenarios. It is found that KAT-LM[UM] fluctuates more often before 1000 iterations, which might be a consequence of larger gradient noises.

\begin{table}[t] 
	\caption{Comparison of Accuracy Performance on Test Data}
	\label{tab:acc} 
	\setlength\tabcolsep{10pt}  
	\begin{threeparttable} 
		\begin{tabular}{lccc}
			\toprule
			Scenario   & RMSE~(MW) & MAPE~[\%] & sMAPE~[\%] \\
			\midrule
			KAT-LM[UM] & \textbf{0.693}     & \textbf{15.1}      & \textbf{11.5}       \\
			O-LM[UM]   & 1.407     & 21.4      & 19.7       \\
			O-SM[UM]   & 1.534     & 23.7      & 22.2       \\
			O-RM[UM]   & 1.756     & 34.0      & 25.6       \\
			DA1-LM[UM] & 1.491     & 26.9      & 21.2       \\
			DA2-LM[UM] & 1.274     & 19.9      & 18.3       \\
			\bottomrule
		\end{tabular}
		\begin{tablenotes}
			\item Note: ``sMAPE'' stands for symmetric mean absolute percentage errors. The most accurate items in each column are highlighted.
		\end{tablenotes}
	\end{threeparttable}
\end{table} 

Table~\ref{tab:acc} takes a close look at the accuracy performance by three popular metrics: root mean squared error (RMSE), the original and symmetric versions of mean absolute percentage error (MAPE and sMAPE). It is not surprising that KAT-LM[UM] still makes the best performances in all metrics, with an average drop of 53.6\% in RMSE, 40.0\% in MAPE, and 46.3\% in sMAPE. Although modeling a single user is often a big challenge, it is possible to reduce the MAPE and sMAPE of KAT-LM[UM] to below 7\% when dropping 10\% low-level consumption data, and KAT-LM[UM] still keeps its superiority in this case. 
Another interesting finding in Table~\ref{tab:acc} is that O-LM[UM] does not perform too bad, which might be related to the inherent regularization effects of using ReLU functions.

\begin{figure}[t]
	\centering
	\includegraphics[width=0.42\textwidth]{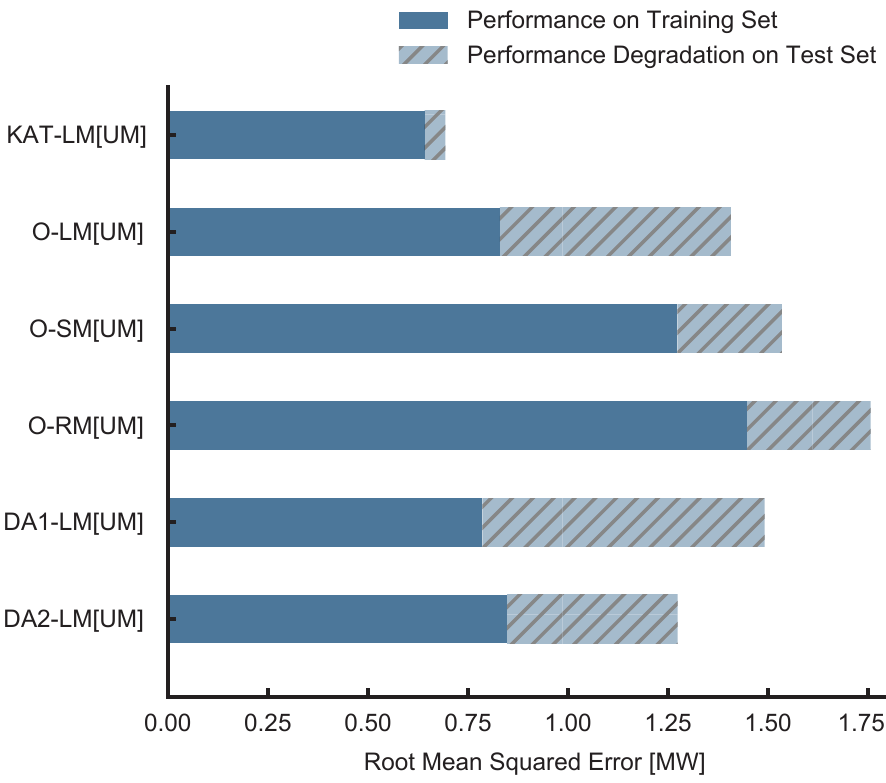}
	\caption{Analysis of the overfitting observations of different scenarios. The difference between training and test performance is shown in the hatched bars for quick comparison.}
	\label{fig:overfitting}
\end{figure}

When analyzing the results in Fig.~\ref{fig:loss-series} and Table~\ref{tab:acc} together, the diverse performances on training (Fig.~\ref{fig:loss-series}) and test datasets (Table~\ref{tab:acc}) are discernible. Fig.~\ref{fig:overfitting} reveals such overfitting effects in a clear way. KAT-LM[UM] keeps the lowest error level (0.668 on average) while this performance degradation on test dataset is mild (only 0.052). On the contrary, overfitting can be found more or less in other scenarios, especially DA1-LM[UM] and O-LM[UM] (0.705 and 0.578 degradation). From these facts, the proposed framework and methodology has truly helped increase the generalization performance.

\subsubsection{Method Effectiveness}
Above results have validated the superior performance of KAT-LM[UM] from the perspective of training efficiency (training speed, convergence value) and training quality (overfitting). We next turn to further explain the effectiveness by more statistical and visualization techniques. Results below only consider O-LM[UM] and KAT[UM] to avoid multiple influential factors.

\begin{figure}[t]
	\centering
	\includegraphics[width=0.5\textwidth]{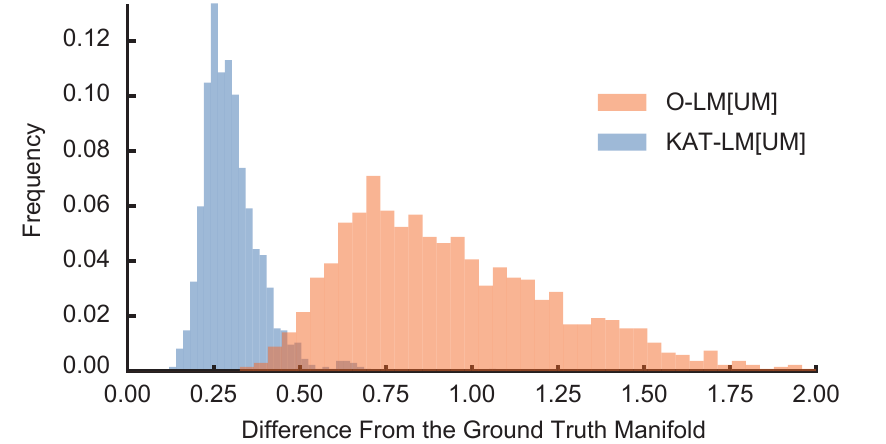}
	\caption{Distribution of prediction errors of O-LM[UM] and KAT-LM[UM] on each synthetic data position. Some  larger but less frequent data may locate beyond the visible region for better visualization.}
	\label{fig:manifold-diff}
\end{figure}

Fig.~\ref{fig:manifold-diff} is focused on the impacts of those inaccurate synthetic data. Following the analysis from Subsection~\ref{subsec:prediction-error}, we calculate the prediction errors of O-LM[UM] and KAT-LM[UM] at each point where a synthetic data is located. In Fig.~\ref{fig:manifold-diff}, a tall-and-skinny distribution is found in KAT-LM[UM], indicating a small average error that is only 32.3\% of that in O-LM[UM]. This validates that those inaccurate synthetic data are able to provide imperfect but useful manifold information.

\begin{figure}[t]
	\centering
	\includegraphics[width=0.5\textwidth]{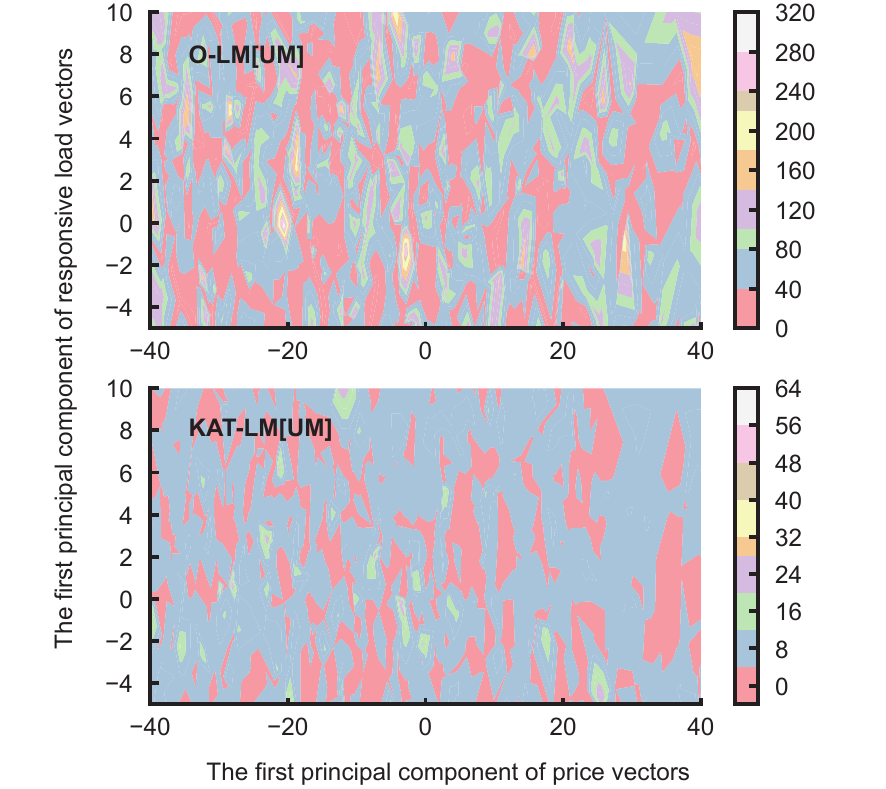}
	\caption{Comparison of the approximation errors of loss functions. We only focus on O-LM[UM] and KAT-LM[UM], and the price and responsive electricity consumption vectors are processed by principal component analysis to make a contour map. Best viewed in color.}
	\label{fig:loss-landscape}
\end{figure}

Fig.~\ref{fig:loss-landscape} analyzes the loss functions by visualizing the approximation errors. Following the statements from Subsection~\ref{subsec:loss-func}, we calculate the errors between ground truth loss function and the loss functions of O-LM[UM] and KAT-LM[UM]. For visualization, the price vectors and responsive electricity consumption vectors are reduced to one dimension by conducting principal component analysis (PCA) and only keeping the first component, and the function difference can then be visualized in Fig.~\ref{fig:loss-landscape}. It is clear that the overall approximation errors of KAT-LM[UM] are much smaller than O-LM[UM]: KAT-LM[UM] has an average error of 5.688, an order of magnitude smaller than the 55.131 of O-LM[UM]. One may also find that the local landscape of KAT-LM[UM] is smoother, which often implies the better generalization potential.

The last focus of this case is on the gradient noise during training (4000 iterations in total). From a statistical perspective, the average gradient noise is 308.776 for KAT-LM[UM], but only 23.481 for O-LM[UM]. Also, the maximal gradient noises of KAT-LM[UM] is 12.7 times as much as that of O-LM[UM]. Rich gradient noises could generally enhance the local search capability of KAT-LM[UM] and contribute to finding a better set of model parameters.

\subsection{Case 2: Probabilistic Price Forecasting}

\subsubsection{Description}
Price forecasting is an essential task for market participants, especially in bidding and risk management. Deep learning models as well as probabilistic forecasting are widely applied in this area because electricity prices are highly volatile in nature, and therefore hard to predict.

We collect data from CAISO market, including the real-time prices, electricity consumption, temperature observations, generation of wind, solar, and gas-fired generators. Hourly data spanning over three years, from January 2017 to March 2019, are considered in this case. We typically focus on the probabilistic price forecasting in the first quarter of 2019, and due to the diverse patterns in different quarters, we only pick the first-quarter data from 2017 and 2018 for training. In total, 4320 data are finally selected.

\begin{figure}[t]
	\centering
	\includegraphics[width=0.5\textwidth]{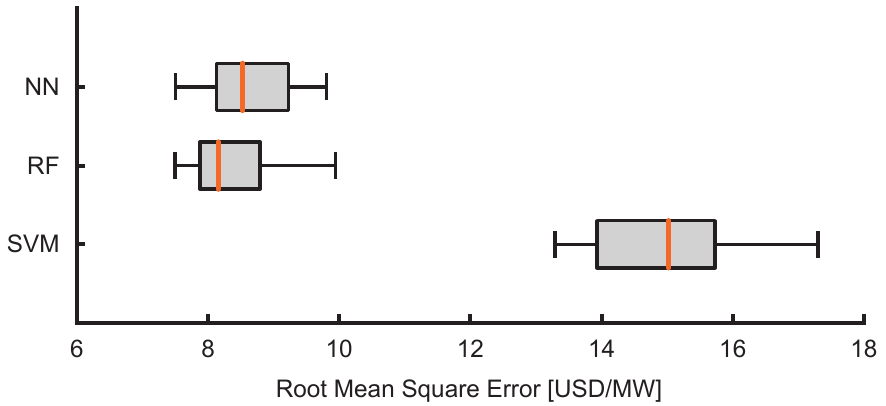}
	\caption{Calibration performance of three kinds of classical models: neural networks, random forests, and support vector machines.}
	\label{fig:cm}
\end{figure}

Three data augmentation techniques will each generate 10000 synthetic data to enlarge the entire dataset. The classical model library involves the support vector machine (SVM), random forest (RF), and dense neural network, all of which have been effective and popular in practice~\cite{lee2018bivariate}. The base settings are shown below: SVM uses the rbf kernel, $\gamma = C = 10$; RF has 50 tree estimators; NN has four hidden layers, each with 30 units. By perturbing the model settings, 10 models of each kind are formulated respectively, and their performance on calibration accuracy is shown in Fig.~\ref{fig:cm}. One may find that SVMs become the worst, and RFs perform slightly better than NNs. We typically set $\alpha=0, \beta=10^2$ in (\ref{eqn:ensemble}) to aggregate all the predictions when making synthetic datasets.

We focus on the application of recurrent neural network (RNN) in probabilistic price forecasting, and extend the case settings in \cite{zeng2019dynamic} to further estimate the 25\%, 50\%, 75\% quantiles. The desired RNN model should have an output of the price prediction, and an input of the real-time price, electricity consumption, temperature, generation of wind, solar, and gas-fired generators within the past three hours. Pinball loss function is used as the loss function, and tanh is the default activation function for every recurrent cell. The detailed structures of LM is 6 cell layers (each with 30 units) plus a dense layer for output. Comparing with the LM structure, SM has 5 cell layers (each with 20 units) instead, and RM adds a dropout rate of 0.2 on each cell layer.

\begin{table}[t] 
	\caption{Comparison of Pinball Loss Values of Different Estimated Quantiles [USD/MW]}
	\label{tab:pinball-loss} 
	\setlength\tabcolsep{10pt}  
	\begin{threeparttable} 
		\begin{tabular}{lcccc}
			\toprule
			Scenario   & QT-25\% & QT-50\% & QT-75\% & Mean \\
			\midrule
			KAT-LM[PPF]   & \textbf{1.952} & \textbf{2.491} & \textbf{2.361} & \textbf{2.268} \\
			O-LM[PPF]     & 2.699 & 2.976 & 3.024 & 2.900 \\
			O-SM[PPF]     & 2.322 & 2.909 & 2.845 & 2.692 \\
			O-RM[PPF]     & 4.206 & 4.922 & 4.663 & 4.597 \\
			DA1-LM[PPF]   & 2.547 & 2.896 & 3.091 & 2.844 \\
			DA2-LM[PPF]   & 2.766 & 3.069 & 2.545 & 2.793 \\
			\bottomrule
		\end{tabular}
		\begin{tablenotes}
			\item Note: ``QT'' stands for quantile. The mean pinball losses of three quantiles are given in the last column, and the most accurate items in each column are highlighted as well.
		\end{tablenotes}
	\end{threeparttable}
\end{table} 

\subsubsection{Accuracy Performance}
In total, 18 RNNs are constructed by combining six scenarios and three quantiles. These RNNs are then trained and tested by the pinball loss values, as shown in Table~\ref{tab:pinball-loss}. KAT-LM[PPF] performs the best in all quantiles and the mean value, while O-RM[PPF] has the worst performance. It seems dropout regularization may have obvious impacts on RNNs' fitting capacity.

\begin{table}[t] 
	\caption{Evaluation of Different Probabilistic Estimation Scenarios}
	\label{tab:prob-score} 
	\begin{threeparttable} 
		\begin{tabular}{lccc}
			\toprule
			Scenario & WS~[USD/MW]     & CRPS~[USD/MW]  & CRPSS~[\%]   \\
			\midrule
			KAT-LM[PPF]   & \textbf{17.287} & \textbf{4.233} & \phantom{-}\textbf{20.3}  \\
			O-LM[PPF]     & 22.894 & 5.315 & \phantom{-0}0.0   \\
			O-SM[PPF]     & 20.669 & 4.985 & \phantom{-0}6.2   \\
			O-RM[PPF]     & 36.470 & 8.581 & -61.5 \\
			DA1-LM[PPF]   & 22.550 & 5.267 & \phantom{-0}0.9   \\
			DA2-LM[PPF]   & 21.474 & 5.031 & \phantom{-0}5.3  \\
			\bottomrule
		\end{tabular}
		\begin{tablenotes}
			\item Note: ``WS'' stands for Winkler score, ``CRPS'' for continuous ranked probability score, and ``CRPSS'' for continuous ranked probability skill score. We take O-LM[PPF] as the baseline for calculating CRPSS values. The most accurate or superior items in each column are also highlighted.
		\end{tablenotes}
	\end{threeparttable}
\end{table} 

Table~\ref{tab:prob-score} will next evaluate the overall performance of these RNNs, using Winkler score (WS), continuous ranked probability score (CRPS), and continuous ranked probability skill score (CRPS). WS is used to assess the 25--75\% uncertainty interval, CRPS will consider all three quantile series, and CRPSS is calculated as the reduction rate of CRPS when taking O-LM[PPF] as the baseline. In Table~\ref{tab:prob-score}, KAT-LM[PPF] has the most accurate performance for all metrics, and in particular, KAT-LM[PPF] has a 20.3\% CRPS improvement (or reduction) than O-LM[PPF].

\begin{figure}[t]
	\centering
	\includegraphics[width=0.40\textwidth]{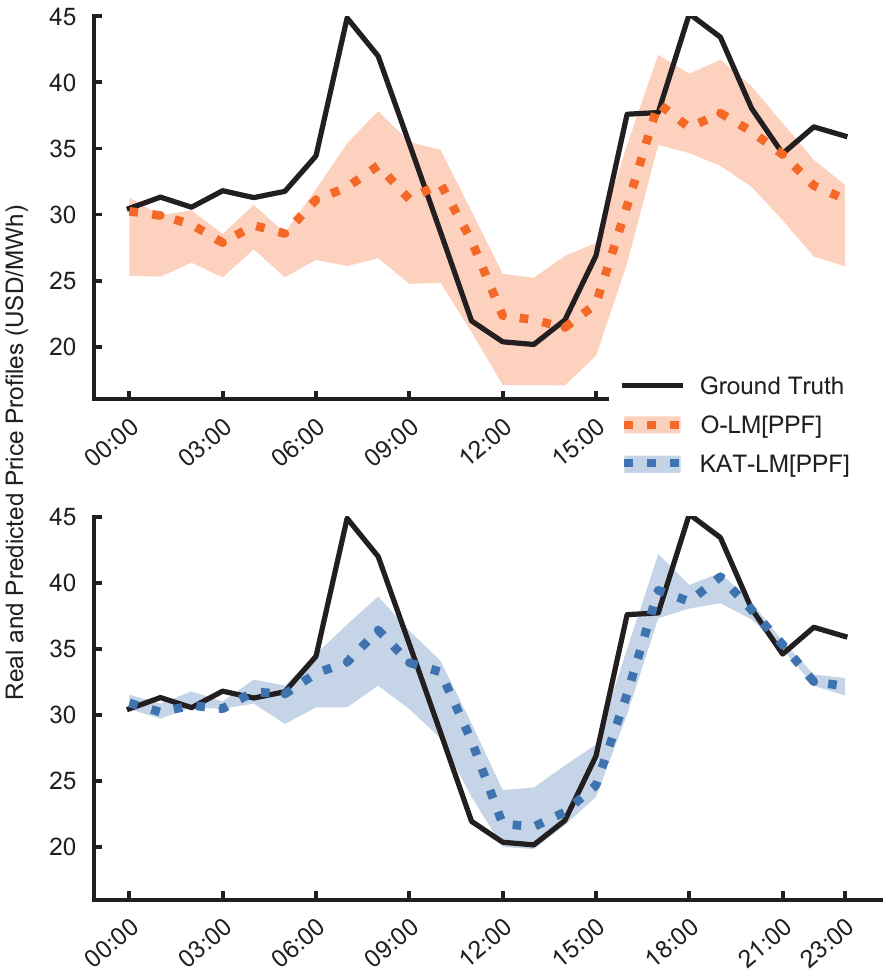}
	\caption{Comparison of probabilistic predictions of O-LM[PPF] and KAT-LM[PPF] on Jan 30, 2019. Given the three quantile profiles (25\%, 50\%, 75\%), this day is typical to demonstrate the KAT-LM[PPF]'s improvements in closer predictions as well as sharper uncertainty intervals.}
	\label{fig:selected-date}
\end{figure}

Fig.~\ref{fig:selected-date} gives a quick preview of the prediction results on one selected date, January 30, 2019. Although both scenarios cannot accurately capture the morning and evening peaks and leave two obvious gaps, KAT-LM[PPF] truly performs better than O-LM[PPF] at least in two aspects: a slightly more accurate predictions (especially before 6AM), and much sharper uncertainty intervals. The forecasting result of KAT-LM[PPF] is thus more informative and valuable for practical use cases.

\begin{table}[t] 
	\caption{Method Validation Under Various Hyper-Parameter Conditions}
	\label{tab:validation} 
	\setlength\tabcolsep{4pt}  
	\begin{threeparttable} 
		\begin{tabular}{ccccc}
			\toprule
			\makecell{Number of\\Cell Layers} & \makecell{Number of\\Cell Units} & \makecell{Activation\\Function}   & \makecell{Reduced Pinball\\Loss~[USD/MW]} & \makecell{Improvement\\Rate~[\%]} \\
			\midrule
			5         & 25      & tanh    & 0.552     & 18.1   \\
			5         & 30      & tanh    & 0.343     & 12.1   \\
			5         & 35      & tanh    & 0.196     & \phantom{0}7.2   \\
			6         & 25      & tanh    & 0.360     & 12.8   \\
			6         & 30      & tanh    & 0.484     & 16.3   \\
			6         & 35      & tanh    & 0.508     & 16.5   \\
			7         & 25      & tanh    & 0.611     & 20.0   \\
			7         & 30      & tanh    & 0.750     & 23.5   \\
			7         & 35      & tanh    & 0.417     & 14.7   \\
			5         & 25      & sigmoid & 0.283     & \phantom{0}9.6   \\
			5         & 30      & sigmoid & 0.500     & 16.8   \\
			5         & 35      & sigmoid & 0.549     & 18.1   \\
			6         & 25      & sigmoid & 0.480     & 15.8   \\
			6         & 30      & sigmoid & 0.555     & 18.1   \\
			6         & 35      & sigmoid & 0.656     & 21.0   \\
			7         & 25      & sigmoid & 0.554     & 17.9   \\
			7         & 30      & sigmoid & 0.506     & 16.6   \\
			7         & 35      & sigmoid & 0.541     & 17.8   \\
			\midrule
			Mean      &         &         & 0.491     & 16.3   \\
			\bottomrule
		\end{tabular}
		\begin{tablenotes}
			\item Note: This table is only focused on O-LM[PPF] and KAT-LM[PPF]. The pinball loss reduction and improvement rate are calculated by taking O-LM[PPF] as the baseline.
		\end{tablenotes}
	\end{threeparttable}
\end{table} 

Table~\ref{tab:validation} conducts a few scanning simulations to test the performance of O-LM[PPF] and KAT-LM[PPF] under various hyper-parameter conditions. This table scans the number of cell layers, cell units, and the selection of activation function. Taking O-LM[PPF] as the baseline, the last two columns of Table~\ref{tab:validation} validate how much pinball loss reduction could be achieved by KAT-LM[PPF], and then calculate the improvement rate. Here our focus is typically on the improvement listed in the fourth column of Table~\ref{tab:validation}. The key message is that KAT-LM[PPF] shows stable improvements under various conditions, and on average, makes a 0.491 value reduction as well as 16.3\% improvement in pinball loss.

\subsubsection{Method Effectiveness}
The method effectiveness can be similarly analyzed and validated as Case~1, and we pay special attention to the gradient noises here, because the recurrent cells in RNNs will result in more complicated gradient features. 

\begin{figure}[t]
	\centering
	\includegraphics[width=0.48\textwidth]{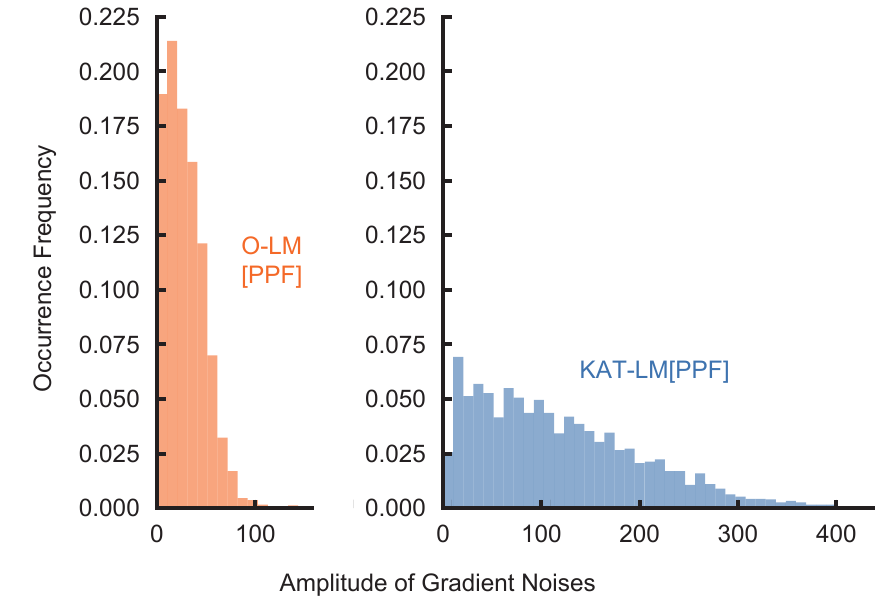}
	\caption{Gradient noise distribution of O-LM[PPF] and KAT-LM[PPF] during training procedure. Quite a few differences in distributional characteristics can be easily validated.}
	\label{fig:grad-noise}
\end{figure}

Fig.~\ref{fig:grad-noise} shows the gradient noise distribution of O-LM[PPF] and KAT-LM[PPF]. Their distributions look quite different, and KAT-LM[PPF] has a much broader distribution, indicating rich noisy observations. In detail, KAT-LM[PPF] has an average gradient noise of 121.342 that is 4.2 times as much as that of O-LM[PPF], and the maximal gradient noise of KAT-LM[PPF] even reaches 742.935. These findings and the accuracy performance in Table~\ref{tab:pinball-loss}--\ref{tab:validation} are compatible with modern machine learning theories presented in Subsection~\ref{subsec:grad-noise}.

\subsection{Extended Tests and Discussions}
We make further efforts in this subsection to validate the proposed framework and models in Case~1 and Case~2. A common concern on the synthetic data is that they are expected to be noisy and may degrade the accuracy performance as a result. Extended tests are simulated below to investigate this issue in Case~1.

The impact of synthetic data can be captured via sensitivity analysis in which we inject additional data noises of different fluctuation rages to imitate various levels of quality. These data noises are randomly generated based on the original synthetic data with $\pm$1--6\% fluctuation, and our focus is on the accuracy performance difference between KAT-LM[UM] and O-LM[UM].

\begin{table}[t] 
	\caption{Sensitivity Analysis on Accuracy Performance Under the Impacts of Noisy Synthetic Data [MW]}
	\label{tab:sensitivity} 
	\setlength\tabcolsep{14.5pt}  
	\begin{threeparttable} 
		\begin{tabular}{cccc}
			\toprule
			Noise    & RMSE of KAT-LM[UM] & Reduced RMSE \\
			\midrule
			$\pm$1.0\%    & 0.795    & 0.612        \\
			$\pm$2.0\%    & 0.862    & 0.545        \\
			$\pm$3.0\%    & 0.897    & 0.510        \\
			$\pm$4.0\%    & 0.898    & 0.509        \\
			$\pm$5.0\%    & 0.931    & 0.476 \\
			$\pm$6.0\%    & 0.974    & 0.433 \\
			\midrule
			Mean     & 0.893    & 0.514  \\
			\bottomrule
		\end{tabular}
		\begin{tablenotes}
			\item Note: The accuracy metric above is RMSE. We calculate the reduced RMSEs in third column by comparing the difference between KAT-LM[UM] and O-LM[UM].
		\end{tablenotes}
	\end{threeparttable}
\end{table} 

Table~\ref{tab:sensitivity} gives the sensitivity results on the model accuracy, which is evaluated by the RMSE metric. One may easily find that the accuracy improvement of KAT-LM[UM], compared with O-LM[UM], will go down when larger noises are injected, reflecting that a high quality synthetic dataset is often useful to boost the learning performance. It is also evident that KAT-LM[UM] still guarantees acceptable performance even when an intermediate level of noises take place in the synthetic data. This finding validates our previous statement that synthetic data are always noisy but still useful in some use case.

Targeting at Case~2, we next formulate two hybrid models from the recent articles~\cite{panapakidis2016day} and \cite{bigerna2018estimating} to further explain the effectiveness of the proposed KAT framework and models. The idea from \cite{panapakidis2016day} is extended to the task of real-time price forecasting where the number of cluster is set to be 2. In this model, namely CT-NN, we still use the same configurations of LM to guarantee a fair comparison. Another model, namely PINN, is inspired by the temperature effect from \cite{bigerna2018estimating} that the heat degree hours and cold degree hours are informative to price forecasting. Based on this finding, the PINN model is established by introducing a physics-based preprocessing step to supplement two extra inputs in LM. Additionally, most scenario settings are kept unchanged, and the major focus is on the accuracy performance. 

\begin{table}[t] 
	\caption{Comparison of Pinball Loss Values and Evaluation Metrics of Different Methods}
	\label{tab:extension} 
	\begin{threeparttable} 
		\begin{tabular}{lccc}
			\toprule
			Metric  & KAT-LM[PPF] & CT-NN[PPF] & PINN[PPF] \\
			\midrule
			PL-25\% [USD/MW] & \textbf{\phantom{0}1.952}  & \phantom{0}2.124  & \phantom{0}2.616         \\
			PL-50\% [USD/MW] & \textbf{\phantom{0}2.491}  & \phantom{0}2.641  & \phantom{0}3.588         \\
			PL-75\% [USD/MW] & \textbf{\phantom{0}2.361}  & \phantom{0}2.379  & \phantom{0}3.187         \\
			WS [USD/MW]      & \textbf{\phantom{}17.287}  & \phantom{}18.012  & \phantom{}23.209        \\
			CRPS [USD/MW]    & \textbf{\phantom{0}4.233}  & \phantom{0}4.313  & \phantom{0}5.733         \\
			CRPSS [\%]       & \textbf{20.4}              & 18.9              & -7.9 \\
			\bottomrule
		\end{tabular}
		\begin{tablenotes}
			\item Note: ``PL-x\%'' stands for the pinball loss at the quantile of x\%. ``WS'' stands for Winkler score, ``CRPS'' for continuous ranked probability score, and ``CRPSS'' for continuous ranked probability skill score. We take O-LM[PPF] as the baseline for calculating CRPSS values. The most accurate or superior items in each row are  highlighted as well.
		\end{tablenotes}
	\end{threeparttable}
\end{table} 

Table~\ref{tab:extension} shows the accuracy performance of KAT-LM[PPF], CT-NN[PPF], and PINN[PPF]. Here all the metrics in Case~2 are taken into consideration. 
The key message is that KAT-LM[PPF] outperforms the other hybrid models in all aspects of interest. Meanwhile, CT-NN[PPF] performs better than PINN[PPF], and approaches KAT-LM[PPF] in some aspects such as the pinball loss at 75\% quantile and the CRPSS.
Another message is that the benefit of hybrid modeling might be uncertain, which closely matches a well-known complaint that these models need case-by-case trial because there is no a general and flexible rule to boost the performance. On the contrary, the proposed KAT framework and models are much more flexible to provide good results.

\section{Conclusion} \label{sec:conclusion}
This paper has proposed a hybrid framework and methodology to take advantage of the domain knowledge and deep learning models. The key techniques include a novel data augmentation technique and an improved training strategy, which are validated to successfully improve the sample efficiency in user modeling and probabilistic price forecasting. We further analyze the underlying reasons and provide explanations for these cases. 

The key findings are twofold: 1) For data-insufficient applications, domain knowledge could largely address the concern of overfitting issues in deep learning. 2) Within the proposed framework, the synthetic data, although imperfect, may still contribute to guiding and calibrating the deep learning models. Taking care of the possible data noises in these synthetic data is critical.

Note that the proposed methodology is rather flexible to cover many more similar applications that may not be discussed in this paper, such as bad data detection, demand response potential estimation, and data privacy. The punchline here is that different model blocks could be integrated in the proposed framework to formulate a lot of combinations, while several trials are needed to finalize the best option.

Our future work will concentrate on the automated and adaptive toolbox that can simplify the configurations of entire process. The current workflow, although powerful, still involves several steps of manual calibrations. In addition, future work will investigate the impacts of synthetic data noises from theoretical and practical perspectives.

\appendices
\section{Nomenclature and Acronyms}

\subsection{Sets, Indices, and Constants}
\begin{IEEEdescription}[\IEEEusemathlabelsep\IEEEsetlabelwidth{$111111$}]
	\item[$\mathit{HD}, \mathit{SD}$] Historical dataset, and synthetic dataset.
	\item[$D$] Hybrid dataset, $D = \mathit{HD} \cup \mathit{SD}$.
	\item[$D_m$] Mini-batch dataset, $D_m \subset D$.
	\item[$i, (i), k$] Indices of data sample, mini-batch data sample, and classical models. 
	\item[$n, m, N$] Size of historical dataset, a mini-batch, and hybrid dataset. Typically $m < n < N$.
	\item[$m_1, m_2$] Split of a mini-batch. $m_1 + m_2 = m$, $m_1 = | D_m \cap \mathit{HD} |$, $m_2 = | D_m \cap \mathit{SD} |$.
	\item[$T, s$] A large number, and the preset step size.
	\item[$\alpha, \beta$] User-defined constants for calculating the weighting factors.
\end{IEEEdescription}

\subsection{Variables}
\begin{IEEEdescription}[\IEEEusemathlabelsep\IEEEsetlabelwidth{$111111$}]
	\item[$x_i, y_i$] Data sample in historical dataset and synthetic dataset. $\{ (x_1, y_1),\cdots,(x_n, y_n) \}$ for historical dataset, while $\{ (x_n, y_n),\cdots,(x_N, y_N) \}$ for synthetic dataset.
	\item[$x_{(i)}, y_{(i)}$] Data sample in a mini-batch.
	\item[$\theta, w_k$] Parameters in a deep learning model and the $k$-th classical model. Typically $\dim w_k < n \ll \dim \theta$.
	\item[$e_k$] Mean squared error between the outputs of $k$-th classical model and actual observations.
	\item[$\gamma_k$] Normalized weighting factor for the $k$-th classical model when aggregating different outputs.
	\item[$\eta$] Proportion of historical data in a mini-batch. $\eta = m_1 / m$. Bounded by the preset minimum $\eta_\textrm{min}$ and maximum $\eta_\textrm{max}$.
\end{IEEEdescription}

\subsection{Functions}
\begin{IEEEdescription}[\IEEEusemathlabelsep\IEEEsetlabelwidth{$111111$}]
	\item[$f(\cdot; \theta)$] Deep learning model, parameterized by $\theta$.
	\item[$g(\cdot; w_k)$] $k$-th classical model, parameterized by $w_k$.
	\item[$h^\text{KKT}(\cdot)$] Compact expression of KKT conditions.
	\item[$I(\cdot)$] Indicator function that outputs 1 when the inner condition is satisfied, and 0 otherwise.
	\item[$J_N(\theta)$] Loss function of the deep learning model.
	\item[$J_m(\theta)$] Approximated loss that is calculated by a mini-batch.
	\item[$J(\theta)$] Ideal loss function in theory.
	\item[$p(x, y)$] Ground truth joint distribution of data sample.
	\item[$\mathit{GN}(\theta)$] Gradient noise function.
\end{IEEEdescription}

\subsection{Acronyms}
\begin{IEEEdescription}[\IEEEusemathlabelsep\IEEEsetlabelwidth{$111111$}]
	\item[KAT] Knowledge-augmented training. This is the proposed framework.
	\item[(R)NN] (Recurrent) neural network.
	\item[KAT-LM] A large-scale deep learning model that is trained by a hybrid dataset. This is the proposed method in case studies.
	\item[O-LM] A large-scale deep learning model that is trained by the original dataset. This is the major baseline in case studies.
	\item[UM] User modeling. This is the Case~1 task.
	\item[PPF] Probabilistic price forecasting. The Case~2 task.
	\item[RMSE] Root mean squared error.
	\item[(s)MAPE] (Symmetric) mean absolute percentage error.
\end{IEEEdescription}

\bibliographystyle{ieeetr}
\bibliography{ref}

\end{document}